%% file: Paper-2013-059.tex
\documentclass[12pt,a4paper]{article}

\usepackage{ifthen} 
\newboolean{pdflatex}
\setboolean{pdflatex}{true} 

\newboolean{articletitles}
\setboolean{articletitles}{true} 

\newboolean{uprightparticles}
\setboolean{uprightparticles}{false} 

\newboolean{inbibliography}
\setboolean{inbibliography}{false} 

\input{preamble}

\begin{document}

\renewcommand{\thefootnote}{\fnsymbol{footnote}}
\setcounter{footnote}{1}

\input{title-LHCb-PAPER}


\renewcommand{\thefootnote}{\arabic{footnote}}
\setcounter{footnote}{0}



\pagestyle{plain} 
\setcounter{page}{1}
\pagenumbering{arabic}


%

\input{alltext}



\input{acknowledgements}



\addcontentsline{toc}{section}{References}
\bibliographystyle{LHCb}
\bibliography{pap_ref}
\end{document}

%% file: preamble.tex
\usepackage{microtype}
\usepackage{lineno}  
\usepackage{xspace} 

\usepackage{graphicx}  
\usepackage{color}
\usepackage{colortbl}
\graphicspath{{./figs/}} 

\usepackage{amsmath} 
\usepackage{amssymb}
\usepackage{amsfonts}
\usepackage{upgreek} 

\newcommand*\patchAmsMathEnvironmentForLineno[1]{%
\expandafter\let\csname old#1\expandafter\endcsname\csname #1\endcsname
\expandafter\let\csname oldend#1\expandafter\endcsname\csname
end#1\endcsname
 \renewenvironment{#1}%
   {\linenomath\csname old#1\endcsname}%
   {\csname oldend#1\endcsname\endlinenomath}%
}
\newcommand*\patchBothAmsMathEnvironmentsForLineno[1]{%
  \patchAmsMathEnvironmentForLineno{#1}%
  \patchAmsMathEnvironmentForLineno{#1*}%
}
\AtBeginDocument{%
\patchBothAmsMathEnvironmentsForLineno{equation}%
\patchBothAmsMathEnvironmentsForLineno{align}%
\patchBothAmsMathEnvironmentsForLineno{flalign}%
\patchBothAmsMathEnvironmentsForLineno{alignat}%
\patchBothAmsMathEnvironmentsForLineno{gather}%
\patchBothAmsMathEnvironmentsForLineno{multline}%
}

\usepackage{hyperref}    
\usepackage[all]{hypcap} 

\input{lhcb-symbols-def} 

\usepackage{cite} 
\usepackage{mciteplus}

%% file: lhcb-symbols-def.tex
\def\lhcb {\mbox{LHCb}\xspace}

\ifthenelse{\boolean{uprightparticles}}%
{

 \def\Pchi        {\ensuremath{\upchi}\xspace}                 
 \def\Ppsi        {\ensuremath{\uppsi}\xspace}

 \def\PDelta      {\ensuremath{\Delta}\xspace}                 
 \def\PXi      {\ensuremath{\Xi}\xspace}                 
 \def\PLambda      {\ensuremath{\Lambda}\xspace}                 
 \def\PSigma      {\ensuremath{\Sigma}\xspace}                 
 \def\POmega      {\ensuremath{\Omega}\xspace}                 
 \def\PUpsilon      {\ensuremath{\Upsilon}\xspace}                 
 
 \def\PB      {\ensuremath{\mathrm{B}}\xspace}                 
                  
 \def\PD      {\ensuremath{\mathrm{D}}\xspace}

 \def\PJ      {\ensuremath{\mathrm{J}}\xspace}                 
 \def\PK      {\ensuremath{\mathrm{K}}\xspace}

 \def\Pb      {\ensuremath{\mathrm{b}}\xspace}                 
 \def\Pc      {\ensuremath{\mathrm{c}}\xspace}

 \def\Pi      {\ensuremath{\mathrm{i}}\xspace}

}
{

 \def\Pchi        {\ensuremath{\chi}\xspace}                 
 \def\Ppsi        {\ensuremath{\psi}\xspace}                 
                  
 \mathchardef\PDelta="7101
 \mathchardef\PXi="7104
 \mathchardef\PLambda="7103
 \mathchardef\PSigma="7106
 \mathchardef\POmega="710A
 \mathchardef\PUpsilon="7107
                  
 \def\PB      {\ensuremath{B}\xspace}                 
                  
 \def\PD      {\ensuremath{D}\xspace}

 \def\PJ      {\ensuremath{J}\xspace}                 
 \def\PK      {\ensuremath{K}\xspace}

 \def\Pb      {\ensuremath{b}\xspace}                 
 \def\Pc      {\ensuremath{c}\xspace}

 \def\Pi      {\ensuremath{i}\xspace}

}

\def\cquark    {\ensuremath{\Pc}\xspace}

\def\bquark    {\ensuremath{\Pb}\xspace}

  \def\Kbar  {\kern 0.2em\overline{\kern -0.2em \PK}{}\xspace}

  \def\Dbar    {\kern 0.2em\overline{\kern -0.2em \PD}{}\xspace}

\def\Bbar    {\ensuremath{\kern 0.18em\overline{\kern -0.18em \PB}{}}\xspace}

\def\jpsi     {\ensuremath{{\PJ\mskip -3mu/\mskip -2mu\Ppsi\mskip 2mu}}\xspace}
\def\psitwos  {\ensuremath{\Ppsi{(2S)}}\xspace}

\def\chiczero {\ensuremath{\Pchi_{\cquark 0}}\xspace}
\def\chicone  {\ensuremath{\Pchi_{\cquark 1}}\xspace}
\def\chictwo  {\ensuremath{\Pchi_{\cquark 2}}\xspace}
  \def\Y#1S{\ensuremath{\PUpsilon{(#1S)}}\xspace}

\def\chic  {\ensuremath{\Pchi_{c}}\xspace}

\def\Lbar {\ensuremath{\kern 0.1em\overline{\kern -0.1em\PLambda}}\xspace}

\def\AT#1     {\ensuremath{A_{\mathrm{T}}^{#1}}\xspace}           

\def\C#1      {\ensuremath{\mathcal{C}_{#1}}\xspace}                       
\def\Cp#1     {\ensuremath{\mathcal{C}_{#1}^{'}}\xspace}                    
\def\Ceff#1   {\ensuremath{\mathcal{C}_{#1}^{\mathrm{(eff)}}}\xspace}        
\def\Cpeff#1  {\ensuremath{\mathcal{C}_{#1}^{'\mathrm{(eff)}}}\xspace}       
\def\Ope#1    {\ensuremath{\mathcal{O}_{#1}}\xspace}                       
\def\Opep#1   {\ensuremath{\mathcal{O}_{#1}^{'}}\xspace}                    

\newcommand{\tev}{\ifthenelse{\boolean{inbibliography}}{\ensuremath{~T\kern -0.05em eV}\xspace}{\ensuremath{\mathrm{\,Te\kern -0.1em V}}\xspace}}
\newcommand{\gev}{\ensuremath{\mathrm{\,Ge\kern -0.1em V}}\xspace}
\newcommand{\mev}{\ensuremath{\mathrm{\,Me\kern -0.1em V}}\xspace}
\newcommand{\kev}{\ensuremath{\mathrm{\,ke\kern -0.1em V}}\xspace}
\newcommand{\ev}{\ensuremath{\mathrm{\,e\kern -0.1em V}}\xspace}
\newcommand{\gevc}{\ensuremath{{\mathrm{\,Ge\kern -0.1em V\!/}c}}\xspace}
\newcommand{\mevc}{\ensuremath{{\mathrm{\,Me\kern -0.1em V\!/}c}}\xspace}
\newcommand{\gevcc}{\ensuremath{{\mathrm{\,Ge\kern -0.1em V\!/}c^2}}\xspace}
\newcommand{\gevgevcccc}{\ensuremath{{\mathrm{\,Ge\kern -0.1em V^2\!/}c^4}}\xspace}
\newcommand{\gevgev}{\ensuremath{{\mathrm{\,Ge\kern -0.1em V^2\!}}}\xspace}
\newcommand{\mevcc}{\ensuremath{{\mathrm{\,Me\kern -0.1em V\!/}c^2}}\xspace}
\newcommand{\ccgevgev}{\ensuremath{{\mathrm{\,Ge\kern -0.1em V^{-2}\!}}}\xspace}
\newcommand{\gevgevcc}{\ensuremath{{\mathrm{\,Ge\kern -0.1em V^2\!/} c^{2}}}\xspace}

\def\mum  {\ensuremath{\,\upmu\rm m}\xspace}

\def\nb {\ensuremath{\rm \,nb}\xspace}

\def\pb {\ensuremath{\rm \,pb}\xspace}
\def\invpb {\ensuremath{\mbox{\,pb}^{-1}}\xspace}

\def\gsim{{~\raise.15em\hbox{$>$}\kern-.85em
          \lower.35em\hbox{$\sim$}~}\xspace}
\def\lsim{{~\raise.15em\hbox{$<$}\kern-.85em
          \lower.35em\hbox{$\sim$}~}\xspace}

\def\pt         {\mbox{$p_{\rm T}$}\xspace}
\def\ptsq         {\mbox{$p_{\rm T}^2$}\xspace}

\def\tell1  {TELL1\xspace}
\def\ukl1   {UKL1\xspace}

\newcommand{\eg}{\mbox{\itshape e.g.}\xspace}

%% file: title-LHCb-PAPER.tex

\begin{titlepage}
\pagenumbering{roman}

\vspace*{-1.5cm}
\centerline{\large EUROPEAN ORGANIZATION FOR NUCLEAR RESEARCH (CERN)}
\vspace*{1.5cm}
\hspace*{-0.5cm}
\begin{tabular*}{\linewidth}{lc@{\extracolsep{\fill}}r}
\ifthenelse{\boolean{pdflatex}}
{\vspace*{-2.7cm}\mbox{\!\!\!\includegraphics[width=.14\textwidth]{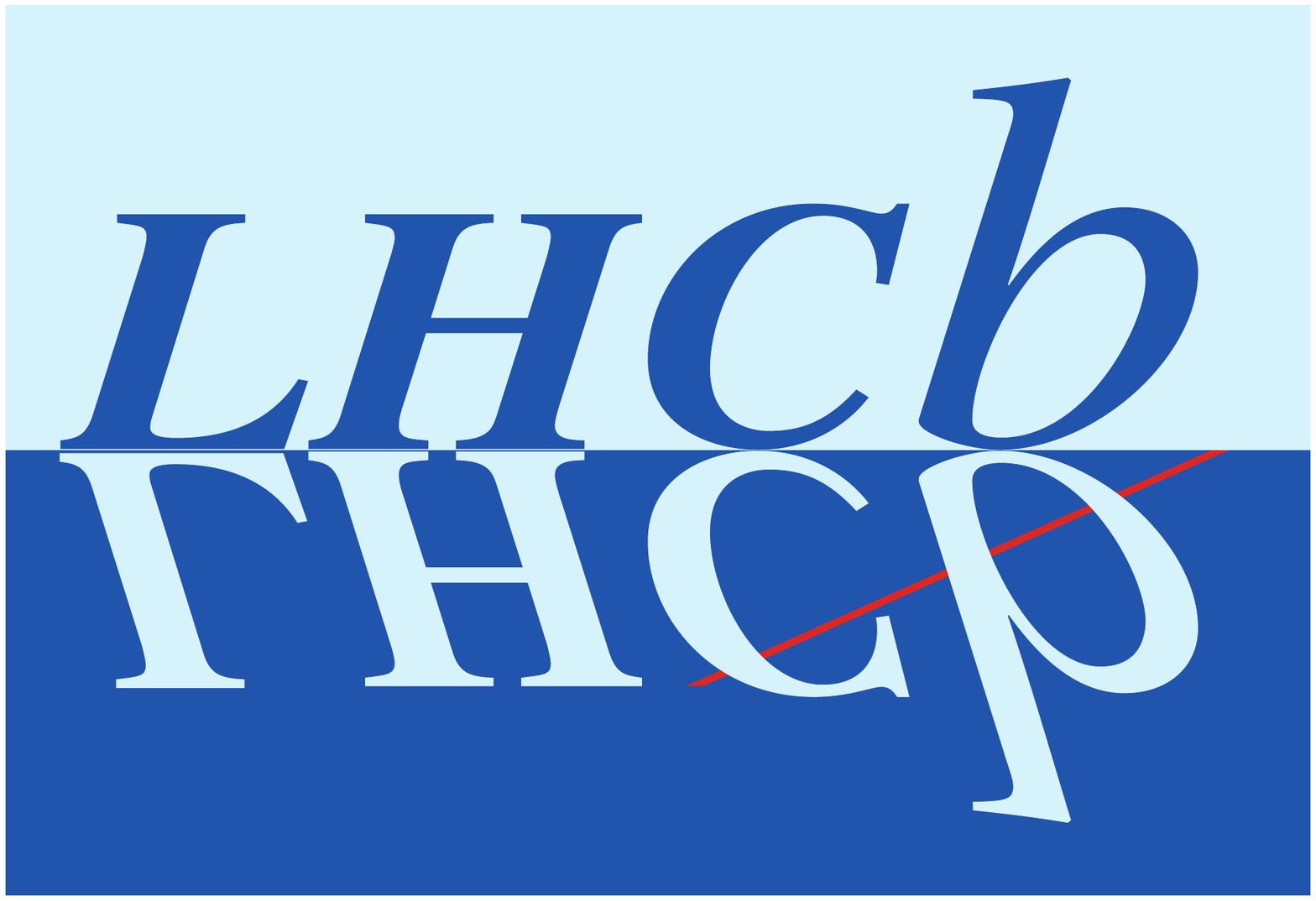}} & &}%
{\vspace*{-1.2cm}\mbox{\!\!\!\includegraphics[width=.12\textwidth]{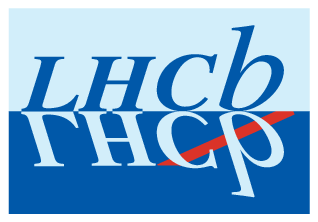}} & &}%
\\
 & & CERN-PH-EP-2013-233 \\  
 & & LHCb-PAPER-2013-059 \\  
 & & 24 January 2014 \\ 
 & & \\
\end{tabular*}

\vspace*{2.0cm}

{\bf\boldmath\huge
\begin{center}
Updated measurements of exclusive 
{\ensuremath{{{\ensuremath{J}\xspace}\mskip -3mu/\mskip -2mu\Ppsi\mskip 2mu}}\xspace}
and $\psi(2S)$
 production cross-sections
in $pp$ collisions at $\sqrt{s}=7$ TeV
\end{center}
}

\vspace*{.5cm}

\begin{center}
The LHCb collaboration\footnote{Authors are listed on the following pages.}
\end{center}

\vspace{\fill}

\begin{abstract}
  \noindent
 The differential cross-section as a function of rapidity has been measured for
the exclusive production of  $J/\psi$ and $\psi(2S)$
 mesons in proton-proton collisions  at \mbox{$\sqrt{s}=7$ TeV},
 using data collected by the LHCb experiment, 
 corresponding to an integrated luminosity of \mbox{930~pb$^{-1}$}.
 The cross-sections times branching fractions to two muons having
pseudorapidities between 2.0 and 4.5 are measured to be
 $$
\begin{array}{rl}
\sigma_{pp\rightarrow\jpsi\rightarrow{\mu^+}{\mu^-}}(2.0<\eta_{\mu^\pm}<4.5)=&291\pm 7\pm19 \pb,\\
\sigma_{pp\rightarrow\psitwos\rightarrow{\mu^+}{\mu^-}}(2.0<\eta_{\mu^\pm}<4.5)=&6.5\pm 0.9\pm 0.4 \pb,\\
\end{array}
$$
where the first uncertainty is statistical and the second is systematic.
The measurements agree with next-to-leading order QCD
predictions as well as with models that include
saturation effects.

\end{abstract}

\vspace*{2.0cm}

\begin{center}
  To be submitted to Journal of Physics, G
\end{center}

\vspace{\fill}

{\footnotesize 
\centerline{\copyright~CERN on behalf of the \lhcb collaboration, license \href{http://creativecommons.org/licenses/by/3.0/}{CC-BY-3.0}.}}
\vspace*{2mm}

\end{titlepage}


\newpage
\setcounter{page}{2}
\mbox{~}
\newpage

\input{LHCb_HD_authorlist_2013-10-10.tex}

\cleardoublepage

%% file: LHCb_HD_authorlist_2013-10-10.tex
\centerline{\large\bf LHCb collaboration}
\begin{flushleft}
\small
R.~Aaij$^{40}$, 
B.~Adeva$^{36}$, 
M.~Adinolfi$^{45}$, 
A.~Affolder$^{51}$, 
Z.~Ajaltouni$^{5}$, 
J.~Albrecht$^{9}$, 
F.~Alessio$^{37}$, 
M.~Alexander$^{50}$, 
S.~Ali$^{40}$, 
G.~Alkhazov$^{29}$
% ...

{\footnotesize \it
% ...
$ ^{5}$Clermont Universit\'{e}, Universit\'{e} Blaise Pascal, CNRS/IN2P3, LPC, Clermont-Ferrand, France\\
% ...
$ ^{9}$Fakult\"{a}t Physik, Technische Universit\"{a}t Dortmund, Dortmund, Germany\\
% ...
$ ^{29}$Petersburg Nuclear Physics Institute (PNPI), Gatchina, Russia\\
% ...
$ ^{36}$Universidad de Santiago de Compostela, Santiago de Compostela, Spain\\
$ ^{37}$European Organization for Nuclear Research (CERN), Geneva, Switzerland\\
% ...
$ ^{40}$Nikhef National Institute for Subatomic Physics, Amsterdam, The Netherlands\\
% ...
$ ^{45}$H.H. Wills Physics Laboratory, University of Bristol, Bristol, United Kingdom\\
% ...
$ ^{50}$School of Physics and Astronomy, University of Glasgow, Glasgow, United Kingdom\\
$ ^{51}$Oliver Lodge Laboratory, University of Liverpool, Liverpool, United Kingdom\\
% ...
}
\end{flushleft}

%% file: alltext.tex
\section{Introduction}
\label{sec:Introduction}

\begin{figure}[b]
\begin{center}
\includegraphics[width=0.9\linewidth]{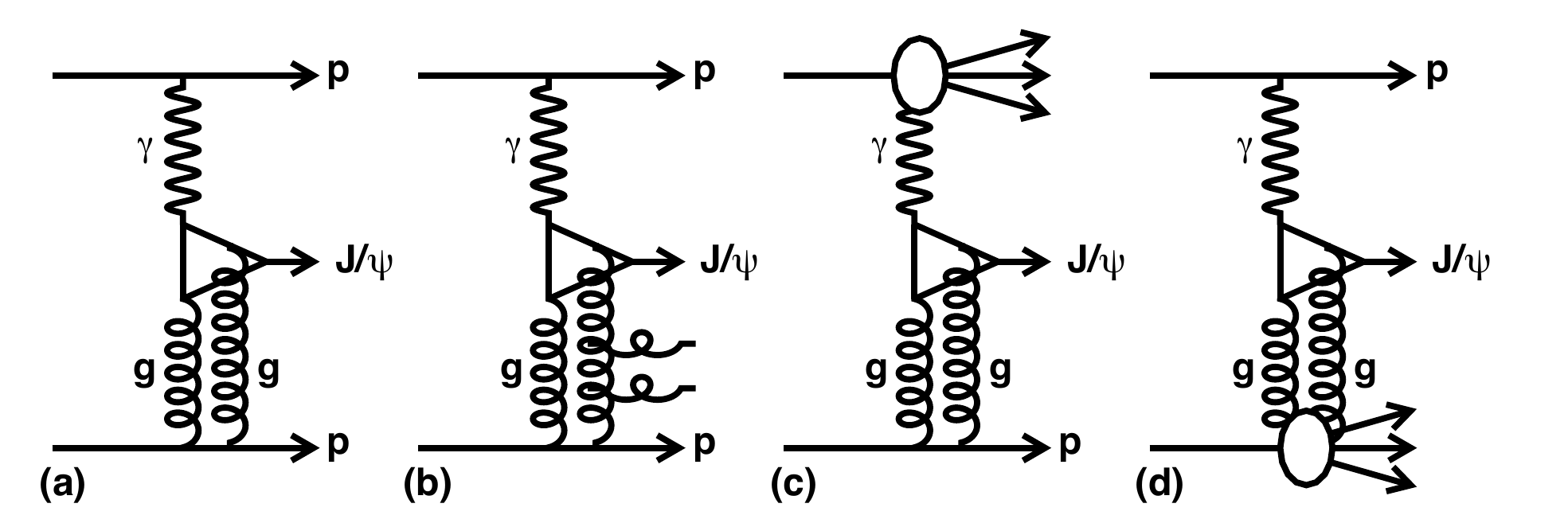}
     \vspace*{-1.0cm}
\end{center}
\caption{
Feynman diagrams displaying (a) exclusive $\jpsi$  and (b) inelastic $\jpsi$ production where a small number of additional particles
are produced due to  gluon radiation and (c),(d) proton dissociation.
Equivalent diagrams apply for \psitwos production.
}
\label{fig:feyn}
\end{figure}

Exclusive \jpsi and \psitwos meson production in hadron collisions are diffractive processes 
that can be calculated in perturbative quantum chromodynamics (QCD)~\cite{ryskin}.  
At leading order they are thought to proceed
via the exchange of a photon and a pomeron, which at sufficiently hard scales can
be described by two gluons  as shown in Fig.~\ref{fig:feyn}(a).
Measurements of exclusive \jpsi and \psitwos 
production thus provide a test of QCD and shed light
on the pomeron, which plays a critical role in the description of diffraction and soft processes.
In particular, the measurements are sensitive to saturation effects~\cite{watt,gay}:
when performed in the pseudorapidity range of the LHCb detector they probe
$x$, the fractional momentum of the parton, down to $5\times 10^{-6}$.
Since the theoretical predictions depend on the gluon parton density function (PDF),
the experimental measurements can be used to constrain it~\cite{alan,jones}. 
Furthermore,
the measurements are sensitive to the presence of the odderon~\cite{odd},
the odd-parity partner of the pomeron, which could mediate the reaction in place of the
photon in Fig.~\ref{fig:feyn}(a).

Several measurements of exclusive \jpsi production have been 
reported by the H1~\cite{h1_jpsi,h1_latest} and
ZEUS~\cite{zeus_jpsi} collaborations at the HERA $ep$ collider, at values of  $W$,
the centre-of-mass energy of the photon-proton system, between 30 and 300\gev.
The first measurement at a hadron machine, at $W\approx80$\gev, 
was made by the CDF collaboration~\cite{cdf} at the Tevatron $p\bar{p}$ collider.
The first measurement in $pp$ collisions was made by the LHCb collaboration~\cite{lhcb}
using data corresponding to an integrated luminosity of $37\invpb$ collected at  $\sqrt{s}=7$\tev, and 
this extended the $W$ reach up to \mbox{1.5\tev}.
Measurements in Pb-Pb collisions at the LHC have been reported by the ALICE collaboration~\cite{alice}.
Measurements of \psitwos production have been 
made by the H1 collaboration~\cite{h1psi}
at four $W$ values  
while both CDF~\cite{cdf} and LHCb~\cite{lhcb} reported results using small samples of \psitwos
consisting of about 40 candidates each.
 
The \jpsi photoproduction cross-section 
has been fit by a power-law function, 
$\sigma_{\gamma p\rightarrow \jpsi p}(W)=a(W/90\gev)^{\delta}$, with the H1 collaboration
measuring
$a=81\pm 3\pb$ and
$\delta=0.67\pm0.03$\cite{h1_latest}.
At leading order this follows from the small-$x$ 
parametrisation of the gluon PDF:
$g(x,Q^2)\propto x^\lambda$ 
at the scale $Q^2=M_\jpsi^2/4$,  
where $M_\jpsi$ is the mass of the \jpsi meson.
All measurements to date at hadron machines are consistent with this,
albeit with rather large uncertainties.
However, higher-order corrections~\cite{jones}
or saturation effects~\cite{watt,gay} 
lead to deviations from a pure power-law behaviour and the measurements presented
here have sufficient precision to probe this effect.
The \psitwos differential cross-section measurements from the H1 collaboration 
are also consistent with
a power-law function, although the limited data sample implies a rather large uncertainty 
and leads to a value for the exponent of $\delta=0.91\pm0.17$~\cite{h1psi}. 
Both CDF and LHCb results are consistent with this.

This paper presents updated measurements from the LHCb collaboration using
\mbox {930\invpb} of data collected in 2011 at $\sqrt{s}=7$ TeV.
Both the \jpsi and \psitwos cross-sections are measured differentially as a function of meson rapidity
and compared to various theoretical models, including those with saturation effects.
The analysis technique is essentially  that published previously~\cite{lhcb}.
The main difference concerns the methodology for determining the background due
to non-exclusive \jpsi and \psitwos production where the additional particles remain undetected.

\section{Detector and data samples}
\label{sec:data}

The \lhcb detector~\cite{lhcbdet} is a single-arm forward
spectrometer covering the \mbox{pseudorapidity} range $2<\eta <5$ (forward region), designed
for the study of particles containing \bquark or \cquark quarks. 
The detector includes a high precision tracking system consisting of a
silicon-strip vertex detector (VELO) surrounding the $pp$ interaction region,
a large-area silicon-strip detector (TT) located upstream of a dipole
magnet with a bending power of about $4{\rm\,Tm}$, and three stations
of silicon-strip detectors (IT) and straw drift-tubes (OT)~\cite{ot} placed
downstream. 
The combined tracking system provides a momentum 
measurement with relative uncertainty 
that varies from 0.4$\%$ at 5$\gevc$ to 0.6$\%$ at 100$\gevc$,
and impact parameter resolution of 20\mum for tracks with large
transverse momentum. 
Different types of charged hadrons are 
distinguished by information from two
ring-imaging Cherenkov detectors~\cite{lhcbrich}. Photon, electron and hadron
candidates are identified by a calorimeter system consisting of
scintillating-pad (SPD) and pre-shower detectors, an electromagnetic
calorimeter and a hadronic calorimeter. The SPD also provides a measurement of the charged particle multiplicity in an event.
Muons are identified by a system composed of alternating layers of iron and multiwire
proportional chambers~\cite{lhcbmuon}. 
The trigger~\cite{lhcbtrig} consists of a hardware stage, based
on information from the calorimeter and muon systems, followed by a
software stage, which applies a full event reconstruction.
The VELO also has sensitivity to charged particles with momenta above $\sim$100$\mevc$ in the \mbox{pseudorapidity} range
 $-3.5<\eta <-1.5$ (backward region), while extending the sensitivity of the forward region to $1.5<\eta<5$.

The $\jpsi$ and $\psitwos$ are identified through their decay to two muons. 
The protons are only marginally deflected by the peripheral collision
and remain undetected inside the beam pipe. 
Therefore, the signature for exclusive vector meson production is an event containing two muons and no other activity. 
Beam-crossings with multiple proton interactions produce additional activity; in the 2011 data-taking
period the average number of visible interactions per bunch crossing was 1.4.
Consequently, requiring an exclusive signature
restricts the analysis to beam crossings with a single $pp$ interaction. 

The \textsc{SuperChic}~\cite{cSUPERC} generator is used to produce samples of 
exclusive $\jpsi$ and $\psitwos$ decays as well as those of the
$\chic$ meson, which form a background for the $\jpsi$ analysis.
These are passed through
a \textsc{GEANT4}~\cite{geant} based detector simulation, the trigger emulation and the event reconstruction chain of the $\lhcb$ experiment.

\section{Event selection}
\label{sec:sel}

The hardware trigger used in this analysis requires a single muon track with transverse momentum $\pt$ $>$ 400$\mevc$ in coincidence with a low SPD multiplicity ($<$ 10 hits).
The software trigger used to select signal events requires two muons with $\pt>400$ \mevc.

\begin{figure}[b]
  \begin{center}
    \includegraphics[width=8cm]{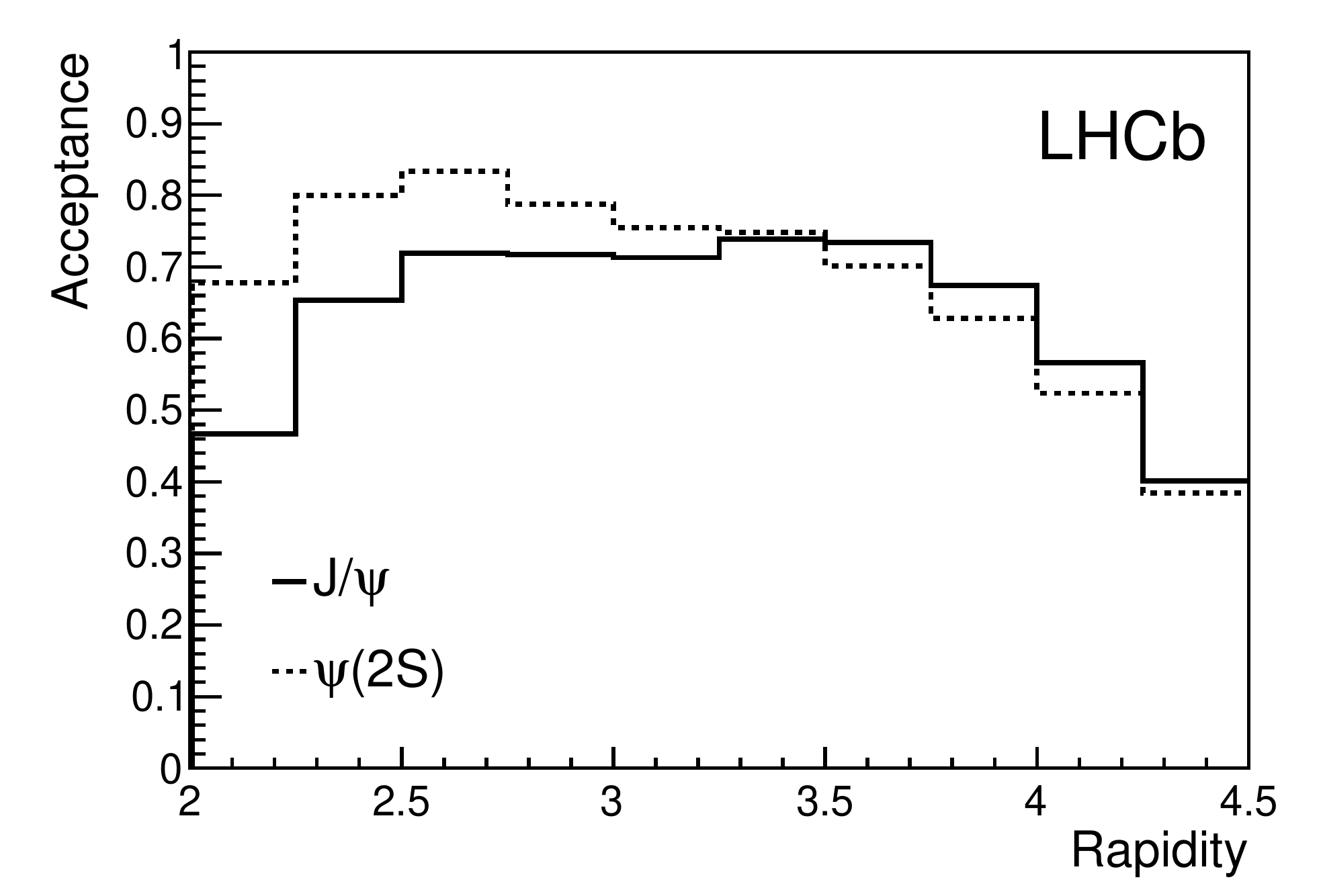}
    \vspace*{-1.0cm}
  \end{center}
  \caption{
    \small 
    Acceptance for (solid) \jpsi and (dotted) \psitwos decays as a function of their rapidity.
       }
  \label{fig:acc}
\end{figure}

The analysis is performed in ten equally sized bins of meson rapidity between 2.0 and 4.5.
The selection of exclusive events begins with the requirement of two reconstructed muons 
in the forward region.
The acceptance of LHCb for muons from \jpsi and \psitwos decays is not uniform: 
muons with low momenta can be swept out of the LHCb acceptance by the magnetic field, or be absorbed before they reach the muon stations.  
Consequently, a fiducial region is defined requiring that each muon has a momentum
greater than 6 \gevc and both tracks are reconstructed within the muon chamber acceptance.
The fiducial acceptance is determined using simulated events 
and is shown in Fig.~\ref{fig:acc}
for both \jpsi and \psitwos decays, both of which are assumed to be transversely polarised
due to $s$-channel helicity conservation. 
A systematic uncertainty of 2\%, fully correlated between bins,
is taken on these values corresponding to the
estimated uncertainty on the description of the tracking in the simulation~\cite{track}.

It is required that there are no photons reconstructed
in the detector and no other tracks that incorporate VELO information.
Unlike the previous analysis~\cite{lhcb}, a veto is not imposed on tracks 
constructed solely from TT and IT/OT detector combinations, as these can 
arise from  detector signals associated with previous beam interactions (``spillover'').
The absence of activity apart from the two muons ensures two rapidity gaps 
which sum to 3.5 units in the forward region.
An additional rapidity gap is obtained by requiring that there are no tracks in the backward region.
The VELO is sensitive to tracks within a certain rapidity region depending on the $z$ position from which the tracks originate and the event topology.  The mean size of the backward rapidity gap that
can be identified is 1.7 with a root mean square of about 0.5.

Muon pairs are combined to form meson candidates whose 
transverse momentum squared must satisfy   
$p_{\rm T}^2<0.8\gevgevcc$,
and whose invariant masses must lie within 65$\mevcc$ of the known $\jpsi$ or $\psitwos$ mass values~\cite{cPDG}.
With these requirements, \mbox{55,985 \jpsi} candidates and 1565 $\psitwos$ candidates are found.  

Three background components are considered: non-resonant background due primarily to
the QED process that produces two muons;  feed-down from exclusive production of
other mesons (\eg $\chi_c$); and inelastic production of mesons where one or
both protons disassociate.

\subsection{Non-resonant background determination}
\begin{figure}[t]
  \begin{center}
        \includegraphics[width=15cm]{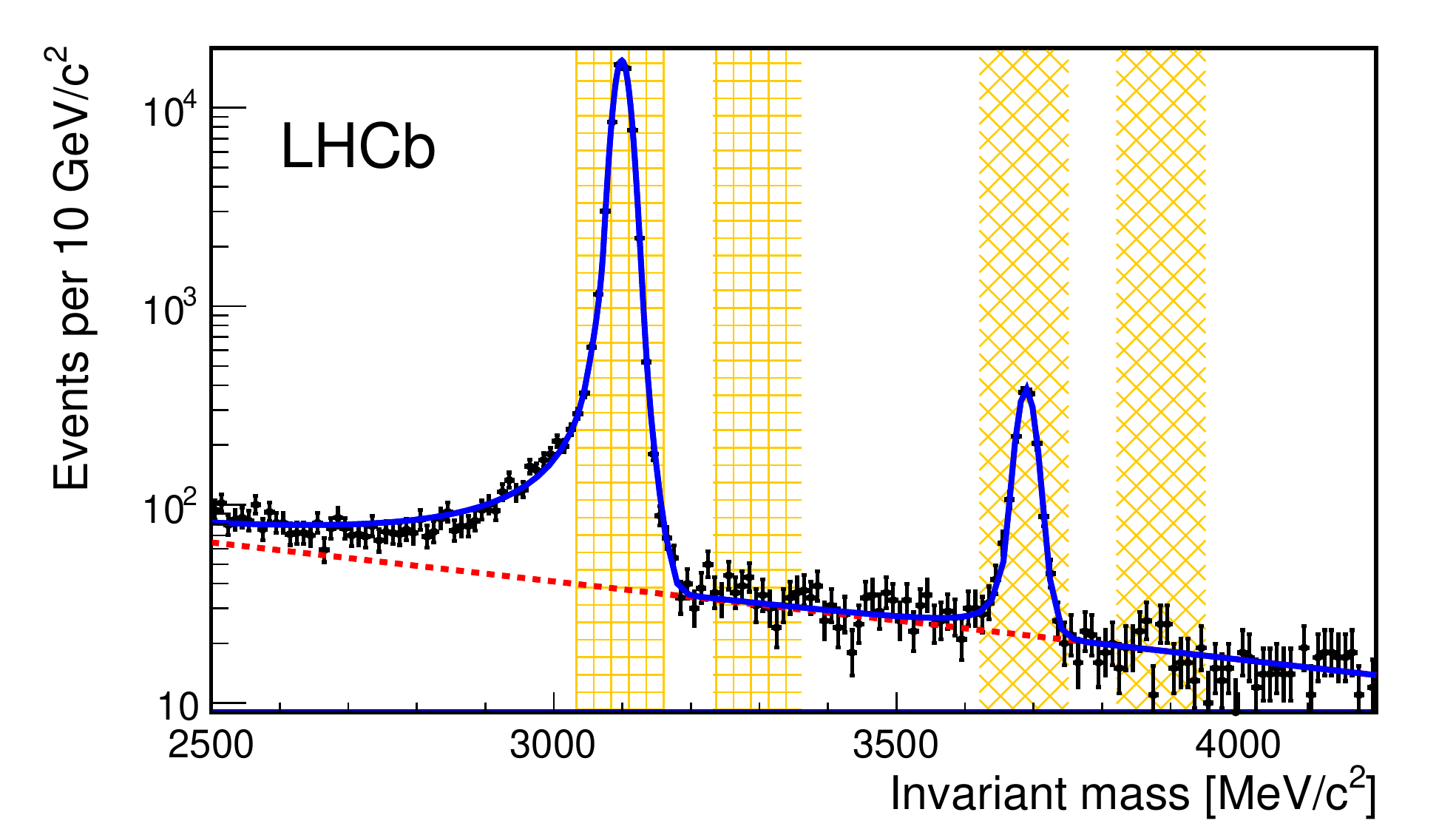}
    \vspace*{-1.0cm}
  \end{center}
  \caption{
    \small 
   Invariant mass distribution of muon pairs after the selection requirements.
The horizontally hatched regions show the \jpsi (left) signal and (right) side-band regions.   
The diagonally hatched regions show the \psitwos (left) signal and (right) side-band regions.   
   The data are fitted (solid curve) with 
   Crystal Ball functions for the signals and an exponential function for the non-resonant background; the latter contribution is shown by the dashed curve.  
   }
  \label{fig:jmass}
\end{figure}

The invariant
mass distributions for \jpsi and \psitwos candidates 
are shown in Fig.~\ref{fig:jmass} and are fitted
with Crystal Ball functions~\cite{Cball} 
to describe the resonant contributions and an exponential function for the non-resonant background.  
Within the range of $\pm 65\mevcc$ about the known meson masses, 
the non-resonant background is estimated to account for  
$(0.8\pm0.1)\%$ 
and $(17.0\pm0.3)\%$ of the \jpsi and \psitwos candidates, respectively.

\subsection{Feed-down  background determination}
\label{sec:feeddown}
Exclusively produced $\chic$ or \psitwos mesons can feed down to 
mimic an exclusive $\jpsi$ decay
when the particles produced in association with the \jpsi remain undetected or
go outside the detector acceptance.
Their contribution is estimated using simulated events normalised to 
an enriched background sample in the data.
Exclusive $\chic$ candidate events are identified in the data as those containing a $\jpsi$ and a single photon~\cite{confnote}. The background from $\chic$ feed-down is then estimated by scaling the number of observed $\chic$ candidates by the ratio of 
simulated \chic mesons passing the \jpsi selection requirements
compared to those identified as \chic candidate events.
The feed-down from $\chic$ decays is estimated to account for \mbox{(7.6 $\pm$ 0.9)$\%$} of the exclusive $\jpsi$ candidates, 
where the uncertainty includes a contribution from the fitted proportions of
\chiczero,\ \chicone,\ \chictwo as well as the photon reconstruction efficiency in simulation.  
Feed-down from $\psitwos$  decays is estimated by scaling 
the \psitwos yields in the resonant peak
(Fig.~\ref{fig:jmass}) by the ratio of 
simulated \psitwos mesons passing the \jpsi selection requirements
compared to those passing the \psitwos selection requirements.
The feed-down from $\psitwos$ decays is estimated to account for (2.5 $\pm$ 0.2)$\%$ of the exclusive $\jpsi$ candidates.

Feed-down into the \psitwos selection is expected to be very small, 
\eg due to $\chi_c(2P)$ or $X(3872)$ decays~\cite{babarx,bellex}.
Relaxing the requirement on the number of photons in the selection,
an additional 2\% of \psitwos candidates is selected, from which a
feed-down of $(2.0 \pm 2.0)$\% is estimated.

\subsection{Inelastic background determination}
The largest background is due to diffractive \jpsi and \psitwos meson production with additional gluon radiation or
proton dissociation (see Figs.~\ref{fig:feyn}(b),(c),(d)) where the particles produced go outside the LHCb acceptance and, in particular, close to the beam-line.
Proton dissociation is more likely to occur when the transverse momentum of the meson is
higher, while additional gluon radiation also increases the average \pt.
Thus the inelastic background has a higher \pt than the
signal.  

In Regge theory, it is assumed that for elastic \jpsi and \psitwos meson production 
\mbox{${d \sigma/ dt}\propto \exp (b_{s}t)$,}
where $t\approx -\ptsq c^2$ is the four-momentum transfer squared at the proton-pomeron vertex.
The H1 and ZEUS collaborations confirmed this dependence and measured 
$b_{\rm s}=4.88\pm0.15\ccgevgev$~\cite{h1_latest}
and 
$b_{\rm s}=4.15\pm0.05^{+0.30}_{-0.18}\ccgevgev$~\cite{zeus_jpsi}, respectively, for \jpsi production,
while H1 measured
$b_{\rm s}=4.3\pm0.6\ccgevgev$~\cite{h1psi} for \psitwos production.
In contrast,  the proton dissociative production of \jpsi or \psitwos
for $|t|<1.2$ \gevgev is observed
to follow
an exponential
dependence, $\exp{(b_{\rm pd}t)}$, 
with $b_{\rm pd}=1.07\pm0.11\ccgevgev$ for \jpsi and $b_{\rm pd}= 0.59\pm0.17\ccgevgev$ for \psitwos~\cite{h1_diffractive}. 
For larger values of $|t|$ a power law is required~\cite{h1_latest}.

The values of $b$ measured at HERA can be extrapolated to LHC energies using
Regge theory:  $b(W)=b_0+4\alpha'\log(W/W_0)$, with $W_0=90 \gev$ and
$\alpha'=0.164\pm 0.041\ccgevgev$~\cite{h1_jpsi} for the elastic process while
$\alpha'=-0.014\pm 0.009\ccgevgev$~\cite{h1_diffractive} for proton dissociation.
This predicts $b_{\rm s}\approx 6\ccgevgev$ and $b_{\rm pd}\approx 1\ccgevgev$ in
the LHCb kinematic region.

After the non-resonant contribution has been subtracted using the
side-bands indicated in Fig.~\ref{fig:jmass},
and with the requirement of $p_{\rm T}^2<0.8\gevgevcc$ for the \jpsi and \psitwos removed, 
the data are fitted to the function
$$
{f_{\rm s}\over N_1}\exp{(-b_{\rm s}\ptsq c^2)}+{ f_{\rm pd}\over N_2}\exp{(-b_{\rm pd}\ptsq c^2)} + 
{f_{\rm fd}\over N_3}F_{\rm fd}(\ptsq),
$$
where
$f_{\rm s}$ and $f_{\rm pd}$ are the fractions of elastic and proton-dissociative production, respectively,
and $f_{\rm fd}$ is the fraction of feed-down
fixed to that obtained in Sec.~\ref{sec:feeddown}.
The shape of the distribution for the feed-down contribution, $F_{\rm fd}$, is 
taken from the data using 
$\chic\rightarrow\jpsi\gamma$ and $\psitwos\rightarrow\jpsi\pi\pi$ candidates.
The numbers
$N_1,N_2$ and $N_3$ normalise each of the three functions to unity in the region $\ptsq<0.8 \gevgevcc$,
while $b_{\rm s}$ and $b_{\rm pd}$ are free parameters.

\begin{figure}[b]
  \begin{center}
    \includegraphics[width=7.5cm]{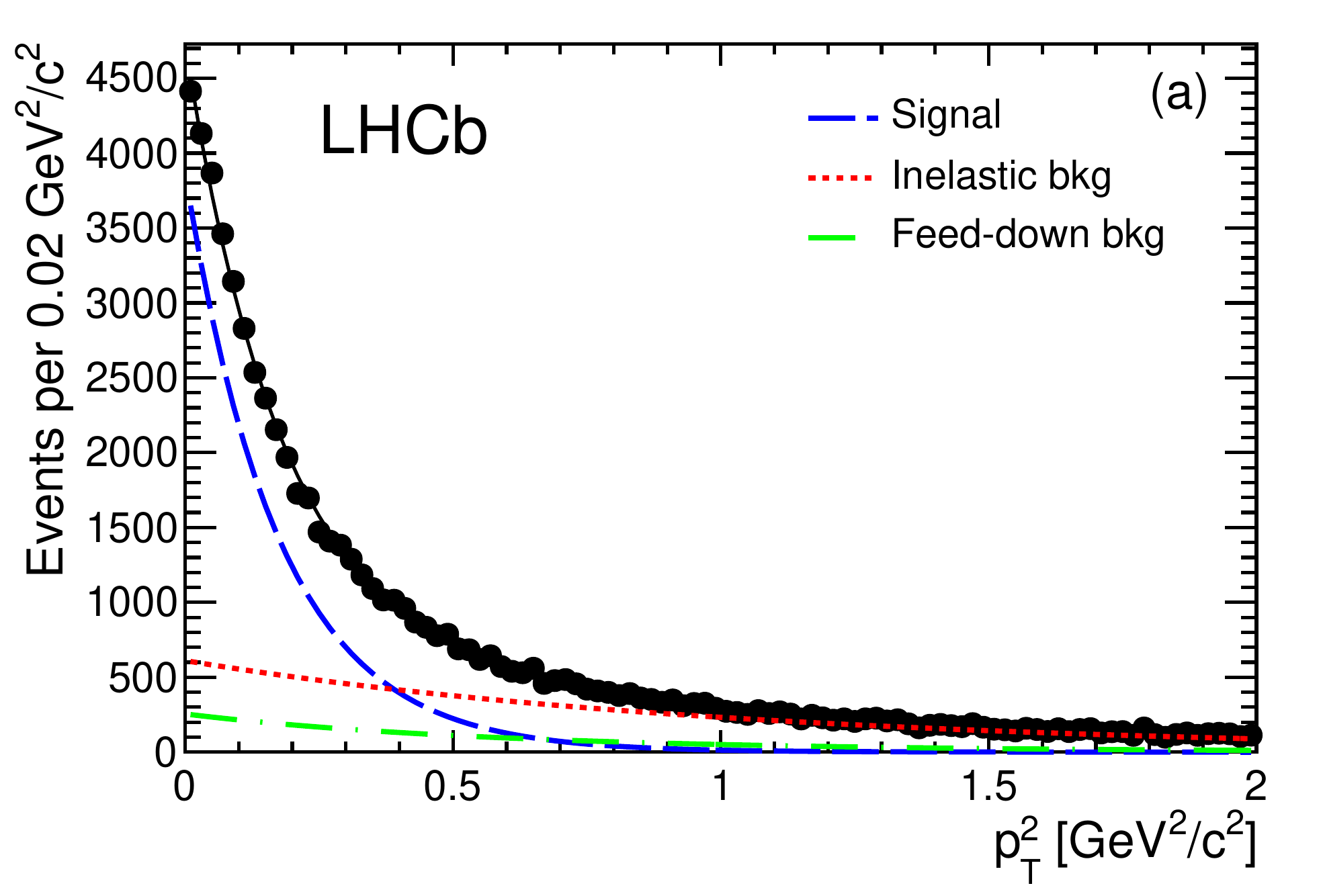}
        \includegraphics[width=7.5cm]{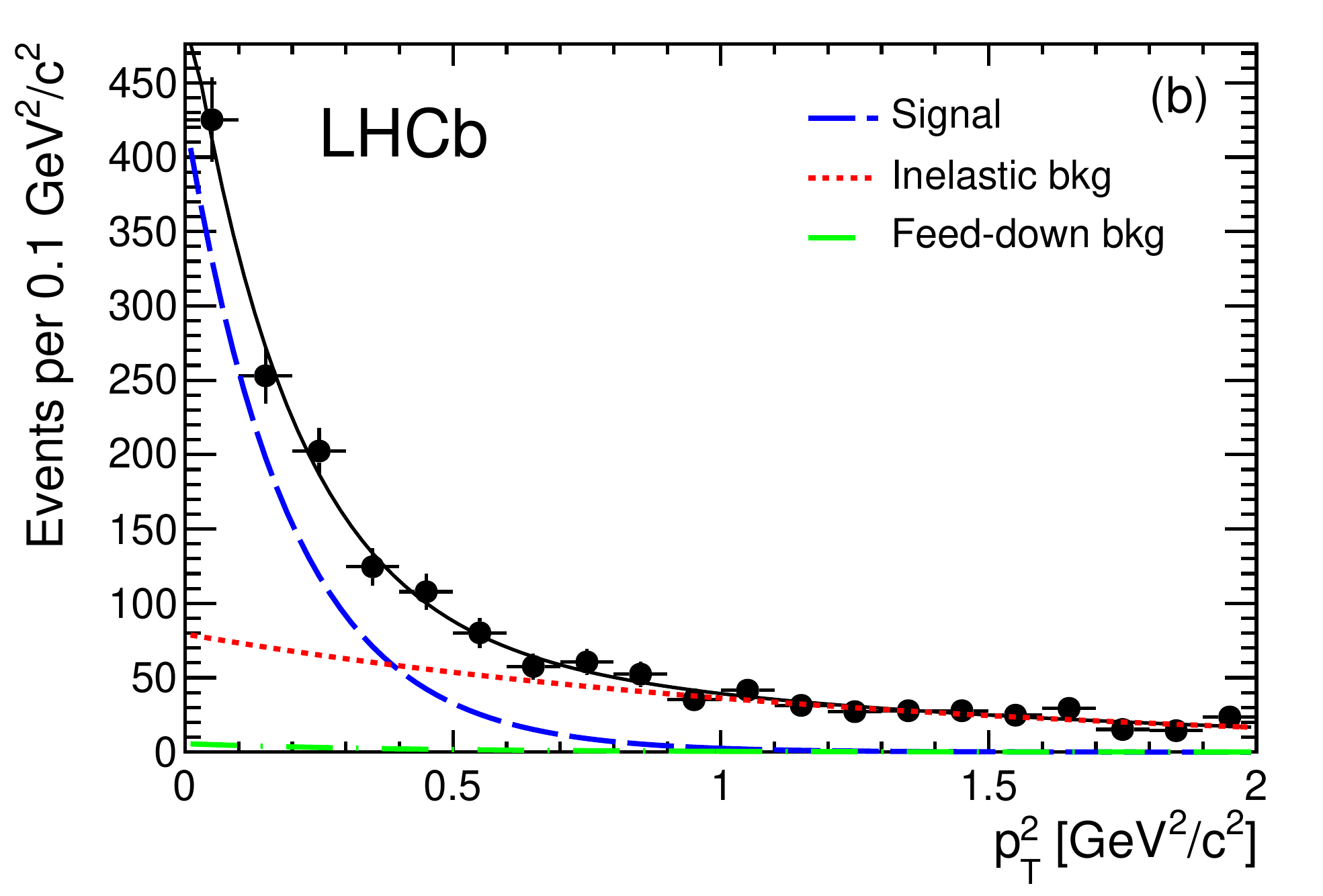}

    \vspace*{-1.0cm}
  \end{center}
  \caption{
    \small 
    Transverse momentum squared distributions for (a) \jpsi and (b) \psitwos candidates,  where the non-resonant background contribution  has been subtracted using side-bands.
    The points are data, the solid curve is the total fit while the different contributions are as described.
    }
  \label{fig:pur}
\end{figure}

The result of the fit for the \jpsi sample is shown in Fig.~\ref{fig:pur}(a).
The $\chi^2/{\rm ndf}$ of the fit is 115/96 and returns values of $b_{\rm s}=5.70\pm0.11\ccgevgev$ and $ b_{\rm pd}=0.97\pm0.04\ccgevgev$.
Below $\ptsq=0.8 \gevgevcc$, the signal fraction is $0.597\pm 0.012$ and correcting for the non-resonant
contribution gives an overall purity for the \jpsi sample of $0.592\pm0.012$. 
The result of the fit for the \psitwos sample is shown in Fig.~\ref{fig:pur}(b).
The $\chi^2/{\rm ndf}$ of the fit is 11/16 and returns values of $b_{\rm s}=5.1\pm0.7\ccgevgev$ and $ b_{\rm pd}=0.8\pm0.2\ccgevgev$.
Below $\ptsq=0.8 \gevgevcc$, the signal fraction is $0.62\pm 0.08$ and correcting for the non-resonant contribution gives an overall purity for the \psitwos sample of $0.52\pm0.07$.
In both cases, the values obtained for $b_{\rm s}$ and $b_{\rm pd}$ are
in agreement 
with the extrapolations of HERA results using Regge theory.

A systematic uncertainty is assigned due to the choice of the fit 
range and the
shape of the parameterisation describing
the inelastic background.
Doubling the range of the fit for the \psitwos candidates changes the signal fraction by 3\%.
Doubling the range of the fit for the \jpsi candidates leads to a poor quality fit; a single exponential function
does not describe the background well.  
For large values of $\ptsq$, the H1 collaboration introduced
a function of the form
$(1+b_{\rm pd}\ptsq/n)^{-n}$ which interpolates between an exponential at low $\ptsq$
and a power law at high $\ptsq$~\cite{h1_latest}. 
Using this functional form and holding $n=3.58$, as determined by H1, gives an acceptable
fit with the signal fraction changing by 5\%.
This value is used as a systematic uncertainty on the purity determination for both 
the \jpsi and \psitwos analyses.

The calculated fraction of feed-down is found to have a negligible effect on the
fraction of signal events as the shapes are similar to the fitted shape
for the proton dissociation.
The purity of the sample is also calculated in bins of rapidity.  No trends are observed and
all values are consistent within statistical uncertainties.  
Consequently a single value for the purity of each sample is assumed, 
independent of the rapidity of the meson.

\section{Cross-section calculation}
\label{sec:cs}

The cross-section times branching fraction to two muons with pseudorapidities between
2.0 and 4.5 
is calculated in ten bins of meson rapidity, $y$, via 
\begin{equation}
\label{eq:cs}
\biggl({d\sigma\over dy}\biggr)_i
=
{\rho N_i\over A_i \epsilon_i \Delta y (\epsilon_{\rm single}L)},
\end{equation}
where:
$N_i$ is the number of events in bin $i$;
$\rho$ is the purity of the sample as described in the previous section;
$A_i$ is the acceptance of the fiducial region as shown in Fig.~\ref{fig:acc}; 
$\epsilon_i$ is the efficiency to select a signal event;
$\Delta y$ is the bin width;
$L$ is the luminosity, which has been determined with an uncertainty of 3.5\%~\cite{lumi};
and $\epsilon_{\rm single}$ is the efficiency for selecting single interaction events, which accounts
for the fact that the selection requirements reject signal events that are accompanied
by a visible proton-proton interaction in the same beam crossing.

The number 
of visible proton-proton interactions per beam crossing, $n$, 
is assumed to follow a Poisson distribution, $P(n)=\mu^{n}\exp(-\mu)/n!$, where $\mu$ is the average number of visible interactions, defined as interactions with one or more tracks having VELO information.
The probability that a signal event is not rejected due to the presence of another visible interaction is given by $P(0)$ and therefore, $\epsilon_{\mathrm{\rm single}}=\exp(-\mu)$.
This has been calculated throughout the data-taking period in roughly hour-long intervals;
variations in $\mu$ during this interval have been studied and found to have a negligible effect.
The spread in the value of $\mu$ for different crossing bunch-pairs is small
and its effect is neglected.
Spillover from previous beam crossings does not affect the VELO, and detector noise is typically at the level of a few uncorrelated
hits, which do not form a track.  
Averaged over the data-taking period, $\epsilon_{\rm single}=0.241\pm0.003$, where the uncertainty
has been calculated with the assumption that at least one track in a visible interaction may be
spurious.

For events with two tracks inside the fiducial region,
the efficiency to select a meson can be expressed as the product
$\epsilon_{\rm id}^\psi\times\epsilon_{\rm trig}^\psi\times\epsilon_{\rm sel}$, where
$\epsilon_{\rm id}^\psi$ is the efficiency for both tracks from the meson to be identified as muons,
$\epsilon_{\rm trig}^\psi$ is the efficiency for an event with two reconstructed muons to fire the triggers,
and $\epsilon_{\rm sel}$ is the efficiency for a triggered event to pass the selection criteria.
Most of these quantities are calculated directly from the data.

For muons coming from \jpsi decays,
a tag-and-probe technique is used to find the efficiency to identify a single muon, 
$\epsilon_{\rm id}^\mu(\phi,\eta)$, as a
function of the azimuthal angle, $\phi$, and pseudorapidity, $\eta$.
The tag is an identified muon that fires the trigger.
The probe is the other track in events with precisely two tracks, which gives the \jpsi
invariant mass when combined with the tag-track.
The fraction of probes identified as muons determines  $\epsilon_{\rm id}^\mu(\phi,\eta)$
from which $\epsilon_{\rm id}^\psi(y)$ is found using the simulation to relate the rapidity
of the meson to the azimuthal angles and pseudorapidities of the two muons.
A scaling factor (typically 1\%), found using simulated events, is applied to get
the corresponding quantity for muons from \psitwos decays, which have a
slightly different momentum spectrum due to the higher mass of the \psitwos meson.

In a similar way, the tag-and-probe technique is used to find the efficiency for a single muon
to fire the hardware trigger, from which is derived 
the corresponding \jpsi efficiency. 
The software trigger requires the presence of two muons.  Its efficiency is determined using
an independent pre-scaled software trigger that fires on a single muon.  Combining these
factors together gives a data-driven estimate of $\epsilon_{\rm trig}^\psi$ for \jpsi decays.  
A correction of typically 3\%, found using simulated events, is applied to get
the corresponding quantity for \psitwos decays.
The values determined for $\epsilon_{\rm id}^\psi\times\epsilon_{\rm trig}^\psi$ as a function of rapidity are 
given in Table~\ref{tab:sum}.

The selection efficiency, $\epsilon_{\rm sel}$, is determined using simulation and data.
From simulation the requirement of having no additional tracks in the event has
an efficiency of $0.997\pm0.001$, and the requirement that $\ptsq<0.8$ \gevgevcc
has an efficiency of $0.979\pm0.003$, where the uncertainty corresponds to a change
of 5\% in the value of $b_{\rm s}$ used in the simulation.
The simulation, calibrated using 
a sample of $\jpsi+\gamma$ candidates in data, determines the efficiency for having
no identified photons in the event to be $0.972\pm 0.005$.  The requirement that the
meson mass lies within 65\mev of the known value is found from the fit in Fig.~\ref{fig:jmass}
and gives an efficiency of $0.96\pm0.01$.
The global requirement, imposed by the hardware trigger, that there be
fewer than ten SPD hits introduces a small inefficiency 
due to  spillover from previous beam crossings.
The efficiency of the SPD requirement is measured to be $0.96\pm0.01$ 
using an independent trigger that has no constraint on
the number of SPD hits.
Combining these together, it is estimated
that $\epsilon_{\rm sel}=0.87\pm0.01$
for the \jpsi selection and the same value is taken for the \psitwos analysis.

The numbers entering the cross-section calculation are summarised in Table~\ref{tab:sum}.
Applying Eq.~\ref{eq:cs} leads to differential cross-section times branching fraction
values for mesons with both
muons inside \mbox{$2.0<\eta_{\mu^\pm}<4.5$}, which are reported in Table~\ref{tab:cs}.
The uncorrelated statistical uncertainties are combined in quadrature and are reported
in the top half of the table. 
The statistical uncertainties on $\epsilon_{\rm sel}$ and the purity are correlated between bins
and are quoted in the lower half of the table.
The systematic uncertainties on the fraction of single interaction beam-crossings, the acceptance,
the shape of the inelastic background and the luminosity are also correlated between bins and are
indicated separately.

The total cross-section times branching fraction to two muons is obtained by 
integrating the cross-sections in Table~\ref{tab:cs}.
The uncorrelated statistical uncertainties are added in quadrature and combined 
with the correlated ones to give the total statistical uncertainty.
The correlated systematic uncertainties, indicated in the lower part of Table~\ref{tab:cs}
 by asterisks, are combined to give the total systematic uncertainty.
 This leads to the following results for the cross-section times branching fraction to
 two muons having pseudorapidities between \mbox{2.0 and 4.5:}
\begin{equation}
\begin{array}{rl}
\sigma_{pp\rightarrow\jpsi\rightarrow{\mu^+}{\mu^-}}(2.0<\eta_{\mu^\pm}<4.5)=&291\pm 7\pm19 \pb,\\
\sigma_{pp\rightarrow\psitwos\rightarrow{\mu^+}{\mu^-}}(2.0<\eta_{\mu^\pm}<4.5)=&6.5\pm 0.9\pm 0.4 \pb,\\
\end{array}
\end{equation}
where the first uncertainty is statistical and the second is systematic.

\begin{table}[t]
\small
  \caption{
    \small 
 Quantities entering the cross-section calculations as a function of meson rapidity.}
\begin{center}\begin{tabular}{lccccc}
    \hline
    $y$ range (\jpsi)     & $[2.00,2.25]$ & [2.25,2.50]  & [2.50,2.75] & [2.75,3.00] & [3.00,3.25] \\
    \hline
    \# Events
    & 798
    & 3911
    & 6632
    & 8600
    & 9987
\\
     Acceptance 
                        & $0.467\pm0.009$ 
                        & $0.653\pm 0.013$ 
                        & $0.719\pm0.014$ 
                        & $0.718\pm 0.014$ 
                        & $0.713\pm0.014$ \\
       $\epsilon_{\rm id}^\psi \times\epsilon_{\rm trig}^\psi $
       & $0.71\pm0.03$ 
       & $0.78\pm0.02$ 
       & $0.81\pm0.01$ 
       & $0.84\pm0.01$ 
       & $0.85\pm0.01$ 
   \\
         Purity  & \multicolumn{5}{c} {$0.592\pm0.012\pm 0.030$} \\		
	  \end{tabular}\end{center}
  
  \begin{center}\begin{tabular}{lccccc}
    \hline
 $y$ range (\jpsi)     & $[3.25,3.50]$ & [3.50,3.75]  & [3.75,4.00] & [4.00,4.25] & [4.25,4.50] \\
    \hline
   
        \# Events & 9877 & 7907 & 5181& 2496 & 596 \\	
        Acceptance  
                & $0.739\pm0.015$ 
        & $0.734\pm0.015$ 
        & $0.674\pm0.014$ 
        & $0.566\pm0.011$ 
        & $0.401\pm0.008$ \\
               $\epsilon_{\rm id}^\psi \times\epsilon_{\rm trig}^\psi $
       & $0.87\pm0.01$ 
       & $0.88\pm0.01$ 
       & $0.87\pm0.01$ 
       & $0.83\pm0.02$ 
       & $0.81\pm0.03$ 
       \\ 
        Purity  & \multicolumn{5}{c} {$0.592\pm0.012\pm0.030$} \\		
  \end{tabular}\end{center}
  
\begin{center}\begin{tabular}{lccccc}
    \hline
$y$ range (\psitwos)     & $[2.00,2.25]$ & [2.25,2.50]  & [2.50,2.75] & [2.75,3.00] & [3.00,3.25] \\
    \hline
    \# Events & 31& 111 & 208 & 1287 & 268 \\
                       Acceptance 
                             & $0.678\pm 0.013$ & $0.800\pm 0.016$ & $0.834\pm0.017$ & $0.787\pm 0.016$ & $0.755\pm0.015$ \\
                       
                         $\epsilon_{\rm id}^\psi \times\epsilon_{\rm trig}^\psi $
       & $0.80\pm0.03$ 
       & $0.83\pm0.02$ 
       & $0.86\pm0.01$ 
       & $0.88\pm0.01$ 
       & $0.88\pm0.01$ 
        \\
        Purity (\psitwos) & \multicolumn{5}{c} {$0.52\pm0.07\pm 0.03$} \\
	  \end{tabular}\end{center}
  
  \begin{center}\begin{tabular}{lccccc}
    \hline
  $y$ range(\psitwos)     & $[3.25,3.50]$ & [3.50,3.75]  & [3.75,4.00] & [4.00,4.25] & [4.25,4.50] \\
    \hline
   
        \# Events&  282   & 201 & 105 & 61 & 11 \\
        Acceptance     & $0.748\pm0.015$ & $0.702\pm0.014$ & $0.628\pm0.013$ & $0.524\pm0.010$ & $0.384\pm0.008$\\
               $\epsilon_{\rm id}^\psi \times\epsilon_{\rm trig}^\psi $
       & $0.90\pm0.01$ 
       & $0.89\pm0.01$ 
       & $0.87\pm0.01$ 
       & $0.84\pm0.02$ 
       & $0.77\pm0.03$ 
	\\	
	        Purity (\psitwos) & \multicolumn{5}{c} {$0.52\pm0.07\pm0.03$} \\	  
	\\
	    \hline
  \multicolumn{2}{l}{$y$ range (\jpsi and \psitwos) }    & \multicolumn{4}{l}{\hspace{2.9cm}$[2.00,4.50] $}\\
    \hline
	               $\epsilon_{\rm sel}$ & \multicolumn{5}{c} {$0.87\pm 0.01$} \\		
                        $\epsilon_{\rm single}$ & \multicolumn{5}{c} {$0.241\pm0.003$} \\		
                $L$ (\invpb)&\multicolumn{5}{c} {$929\pm33$} \\		

  \end{tabular}\end{center}
\label{tab:sum}
\end{table}

\begin{table}[t]
  \caption{
    \small 
  Differential cross-section times branching ratio, in units of \pb, 
   for mesons decaying to two muons, both with $2.0<\eta<4.5$, in bins of meson rapidity. Only the uncorrelated 
  statistical uncertainties are quoted with the central values.  
  The uncertainties, correlated between bins, are tabulated in the lower table.
  Those with an asterisk enter into the systematic uncertainty.
  }
\begin{center}\begin{tabular}{lccccc}
    \hline
    $y$ range     & $[2.00,2.25]$ & [2.25,2.50]  & [2.50,2.75] & [2.75,3.00] & [3.00,3.25] \\
    \hline
    $d\sigma\over dy$ \jpsi  
   & $29.3\pm1.7$
    & $92.5\pm2.4$
    & $137.8\pm2.4$
    & $173.1\pm2.6$
    & $198.0\pm2.7$
                    \\
        $d\sigma\over dy$ \psitwos  
    & $0.56\pm0.11$
    & $1.75\pm0.17$
    & $3.06\pm0.22$
    & $4.41\pm0.26$
    & $4.24\pm0.26$
  \end{tabular}\end{center}
  
  \begin{center}\begin{tabular}{lccccc}
    \hline
        $y$ range     & $[3.25,3.50]$ & [3.50,3.75]  & [3.75,4.00] & [4.00,4.25] & [4.25,4.50] \\
     \hline
        $d\sigma\over dy$ \jpsi  
    & $187.6\pm2.6$
    & $148.9\pm2.4$
    & $107.4\pm2.1$
    & $65.3\pm2.0$
    & $21.9\pm1.3$
\\
        $d\sigma\over dy$ \psitwos  
    & $4.51\pm0.27$
    & $3.43\pm0.25$
    & $2.05\pm0.20$
    & $1.47\pm0.19$
    & $0.36\pm0.11$
  \end{tabular}\end{center}
  
  \begin{center}\begin{tabular}{lc}
    \hline
\multicolumn{2}{c}{Correlated uncertainties expressed as a percentage of the final result} \\
  \hline
  $\epsilon_{\rm sel}$ & 1.4\% \\
  Purity determination (\jpsi) & 2.0\% \\
  Purity determination (\psitwos) & 13.0\% \\
  $^*\epsilon_{\rm single}$ & 1.0\% \\
  $^*$Acceptance & 2.0\% \\
  $^*$Shape of the inelastic background & 5.0\% \\
  $^*$Luminosity & 3.5\% \\
  \hline
  Total correlated statistical uncertainty (\jpsi) & 2.4\% \\
  Total correlated statistical uncertainty (\psitwos) & 13.0\% \\
  Total correlated systematic uncertainty & 6.5\% \\
  \end{tabular}\end{center}

\label{tab:cs}
\end{table}

\begin{table}[h]
  \caption{
    \small 
Fraction of events in a given meson rapidity range where both muons have $2.0<\eta<4.5$.}
\begin{center}\begin{tabular}{lccccc}
    \hline
    $y$ range     & $[2.00,2.25]$ & [2.25,2.50]  & [2.50,2.75] & [2.75,3.00] & [3.00,3.25] \\
    \hline
Acceptance 
& 0.093 & 0.289 & 0.455 & 0.617 & 0.735 \\
  \end{tabular}\end{center}
  
  \begin{center}\begin{tabular}{lccccc}
    \hline
        $y$ range     & $[3.25,3.50]$ & [3.50,3.75]  & [3.75,4.00] & [4.00,4.25] & [4.25,4.50] \\
    \hline
Acceptance 
& 0.738 & 0.624 & 0.470 & 0.286 & 0.103 \\
  \end{tabular}\end{center}
  
\label{tab:acc4pi}
\end{table}

The differential cross-section results reported in Table~\ref{tab:cs} correspond to muons that enter
the LHCb detector fiducial volume.
Differential cross-section results as a function of meson rapidity are obtained by
dividing by the known meson branching fraction to two muons~\cite{cPDG} and by the
fraction of decays with $2.0<\eta_{\mu^\pm}<4.5$.
The former introduces an additional systematic uncertainty of 10\% for the \psitwos measurement.
The latter depends on the kinematics of the decay, is
calculated using {\textsc{SuperChic}} assuming that the \jpsi and \psitwos mesons
are transversely polarised,
and is given in Table~\ref{tab:acc4pi}.

\section{Discussion}

The integrated cross-section measurements for \jpsi and \psitwos mesons
decaying to muons with $2.0<\eta_{\mu^\pm}<4.5$
are compared to various theoretical predictions in Table~\ref{tab:cfth}.
Good agreement is found in each case.

\begin{table}[t]
  \caption{
    \small 
  Comparison of this result to various theoretical predictions.  }
  \begin{center}
  \begin{tabular}{lcc}
    \hline
 & \jpsi [\pb] & \psitwos [\pb]\\
    \hline
    Gon\c{c}alves and Machado~\cite{machado} & 275 & \\
JMRT~\cite{jones} & 282 & 8.3 \\
    Motyka and Watt~\cite{watt} & 334 & \\
        Sch\"afer and Szczurek~\cite{schaefer} & 317 &  \\
    Starlight~\cite{klein04} & 292 & 6.1 \\
    {\textsc{Superchic}}~\cite{cSUPERC} & 317 & 7.0\\
    LHCb measured value & $291\pm 7\pm19$ &  $6.5\pm 0.9\pm 0.4 $ \\

  \end{tabular}\end{center}
  \label{tab:cfth}
\end{table}

\begin{figure}[!htb]
  \begin{center}
    \includegraphics[width=7.5cm]{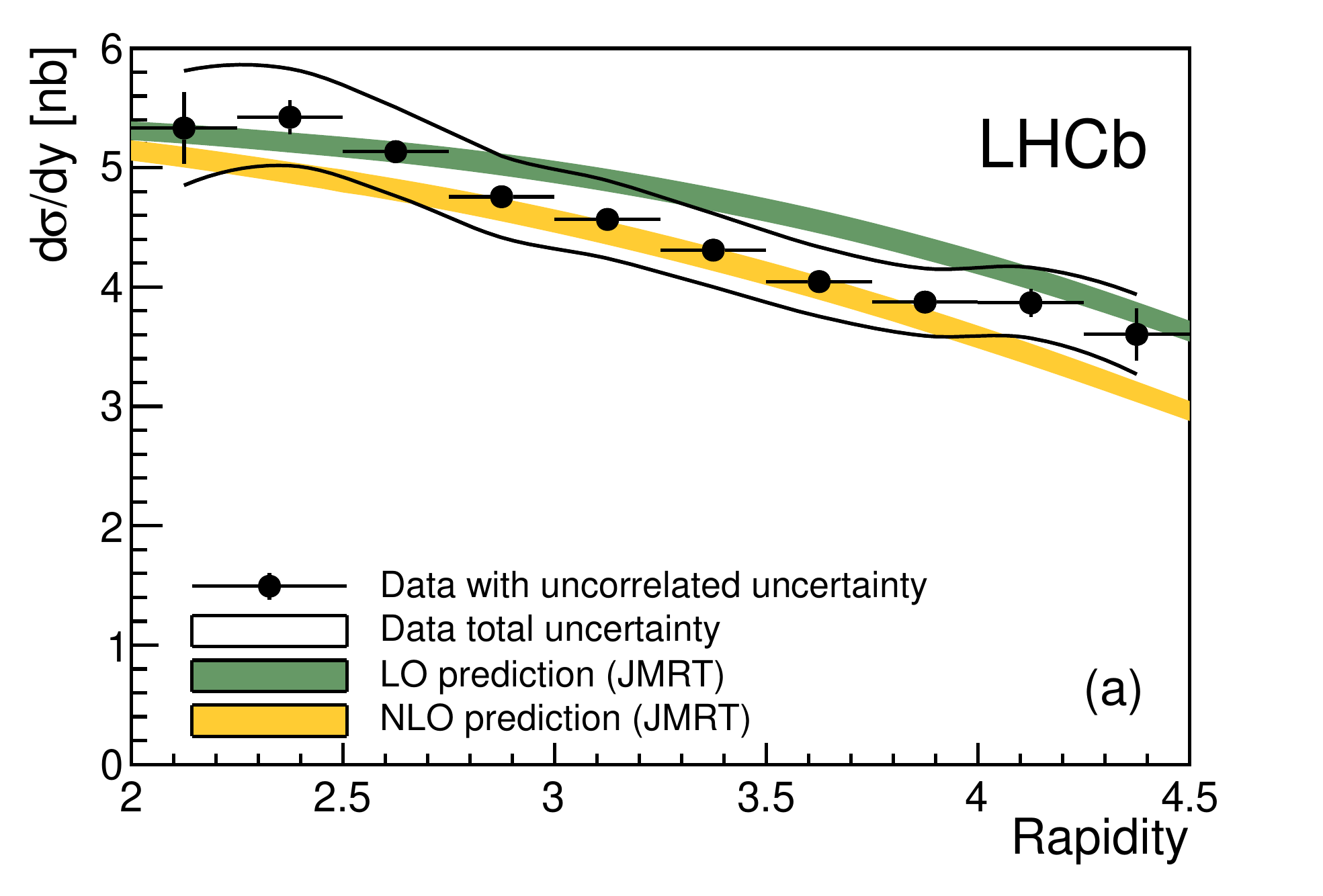}
        \includegraphics[width=7.5cm]{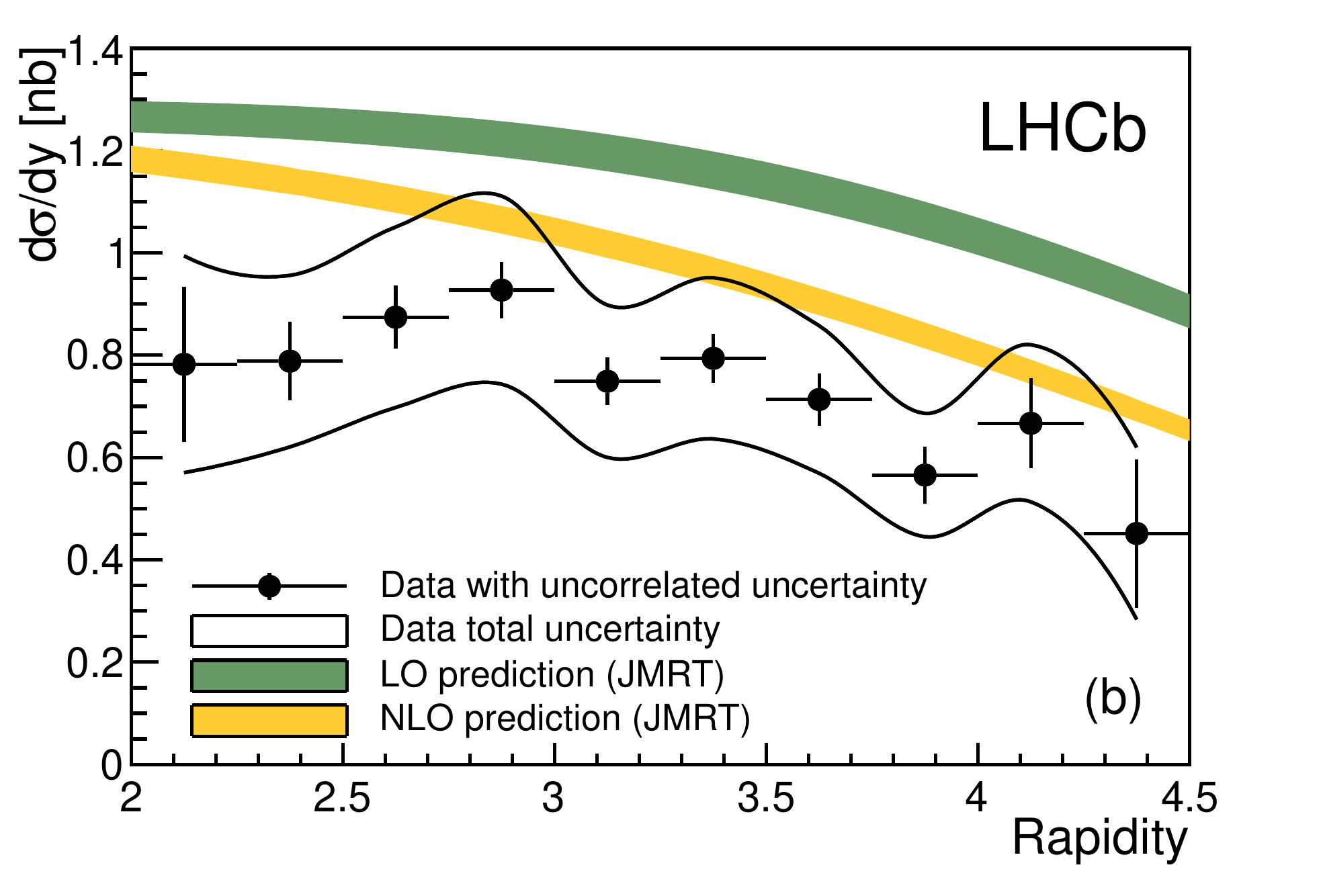}
    \vspace*{-.5cm}
  \end{center}
  \caption{
    \small 
    Differential cross-section for (a)  \jpsi and (b) \psitwos production compared to LO and NLO predictions of \cite{jones}.    The band indicates the total uncertainty, most of which is correlated between bins.
}
  \label{fig:jmrt}
\end{figure}

The differential distribution for \jpsi production 
is presented in Fig.~\ref{fig:jmrt}(a), where the extent of the 
error bars indicates the uncorrelated statistical uncertainties and the band is the total
uncertainty.  
Jones, Martin, Ryskin and Teubner (JMRT)~\cite{jones} have obtained
leading-order (LO) and next-to-leading-order~\footnote{Only the dominant NLO corrections have been considered: see \cite{jones} for details.}
 (NLO) predictions
from a fit to HERA and LHCb 2010 data, which are dominated by the HERA data; 
thus these 
curves can be considered as LO and NLO extrapolations from HERA energies.
The LO result is essentially the power-law photoproduction result from HERA, 
combined with a photon flux function and a gap survival factor~\cite{gap}.
Better agreement is obtained between data and the NLO prediction
than between data and the LO prediction.

\begin{figure}[tb]
  \begin{center}
    \includegraphics[width=12cm]{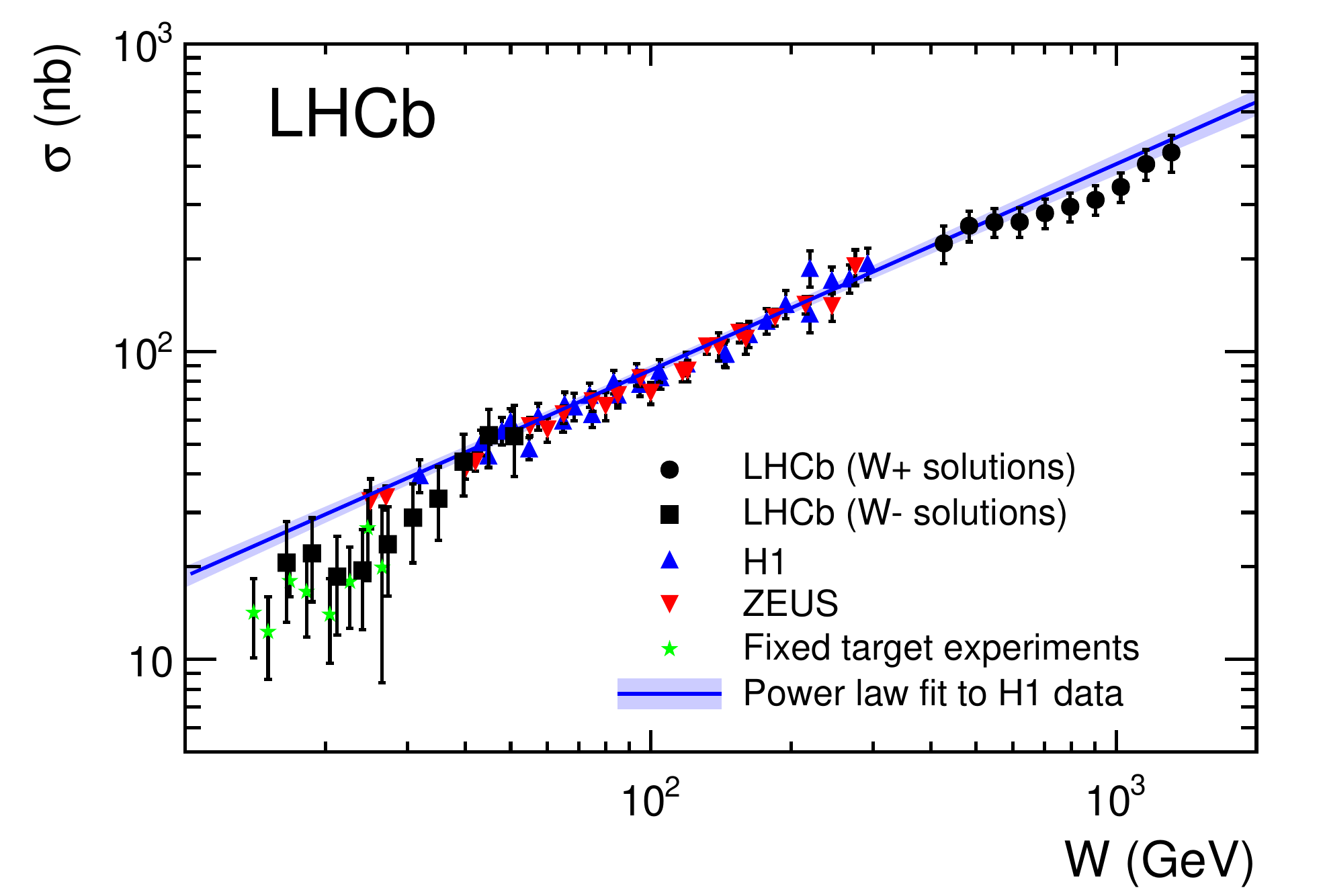}
    \vspace*{-1.0cm}
  \end{center}
  \caption{
    \small 
Photoproduction cross-section as a function of the centre-of-mass of the photon-proton system
with the power-law fit from \cite{h1_latest} superimposed.
The LHCb data points for $W_+(W_-)$ are derived assuming the power-law fit for
$W_-(W_+)$.  The uncertainties are correlated between bins.  Fixed target results are from
the E401~\cite{e401}, E516~\cite{e516} and E687~\cite{e687} collaborations.
       }
  \label{fig:heralhcb}
\end{figure}

Exclusive production of \jpsi in $pp$ collisions is related to photoproduction through
\begin{equation}
{{d}\sigma\over {d}y}_{pp\rightarrow p\jpsi p}=
r_+
k_+{dn\over dk_+}\sigma_{\gamma p\rightarrow \jpsi p}(W_+)
+r_-
k_-{dn\over dk_-}\sigma_{\gamma p\rightarrow \jpsi p}(W_-)
\label{eq:heralhcb}
\end{equation}
where $dn/dk_\pm$ are photon fluxes for photons of energy 
$
k_\pm\approx (M_{\jpsi}/2) \exp(\pm |y|), 
$
\mbox{$(W_\pm)^2=2k_\pm\sqrt{s}$}, and $r_\pm$ are absorptive corrections 
as given, for example, in \cite{schaefer,jones}.
The LHCb results cannot unambiguously determine the photoproduction cross-section
due to contributions from both $W_+$ and $W_-$, corresponding to the 
photon being either an emitter or a target, respectively.  However, a comparison can be made to the
HERA photoproduction results 
using the power-law relationship, 
$\sigma_{\gamma p\rightarrow \jpsi p}(W)=81(W/90\gev)^{0.67}\nb$,
determined by the H1 collaboration~\cite{h1_latest}.
A model-dependent measurement of 
$\sigma_{\gamma p\rightarrow \jpsi p}(W_+)$ 
is obtained from the LHCb differential cross-section measurement 
by
applying Eq.~\ref{eq:heralhcb} and 
assuming the power-law result for
$\sigma_{\gamma p\rightarrow \jpsi p}(W_-)$,
while 
$\sigma_{\gamma p\rightarrow \jpsi p}(W_-)$ is obtained by
assuming the power-law result for 
$\sigma_{\gamma p\rightarrow \jpsi p}(W_+)$.
The result of this procedure is shown in Fig.~\ref{fig:heralhcb}, which compares the
modified LHCb data with HERA and fixed target photoproduction results: 
note that there are two correlated
points plotted for each LHCb measurement, corresponding to the $W_+$ and $W_-$ solutions.
It was shown in our previous publication~\cite{lhcb} that the LHCb data were
consistent, within large statistical uncertainties, with a simple power-law extrapolation of
HERA \jpsi photoproduction results to LHC energies.  
With increased statistics, an extrapolation of the power-law obtained in \cite{h1_latest}
is in marginal agreement with the LHCb data.

The differential distribution for \psitwos production is presented in Fig.~\ref{fig:jmrt}(b) and is
compared to both LO and NLO predictions~\footnote{
These predictions were made by replacing the \jpsi mass and electronic width
by those of the \psitwos neglecting possible relativistic corrections, which may be important for
the heavier meson.
}
from JMRT~\cite{jmrt_psi} 
using the formalism described in \cite{jones} with the gluon PDF taken from their  \jpsi analysis.
Once again, better agreement is found between data and the NLO prediction
than between data and the LO prediction.

\begin{figure}[tb]
  \begin{center}
    \includegraphics[width=7.5cm]{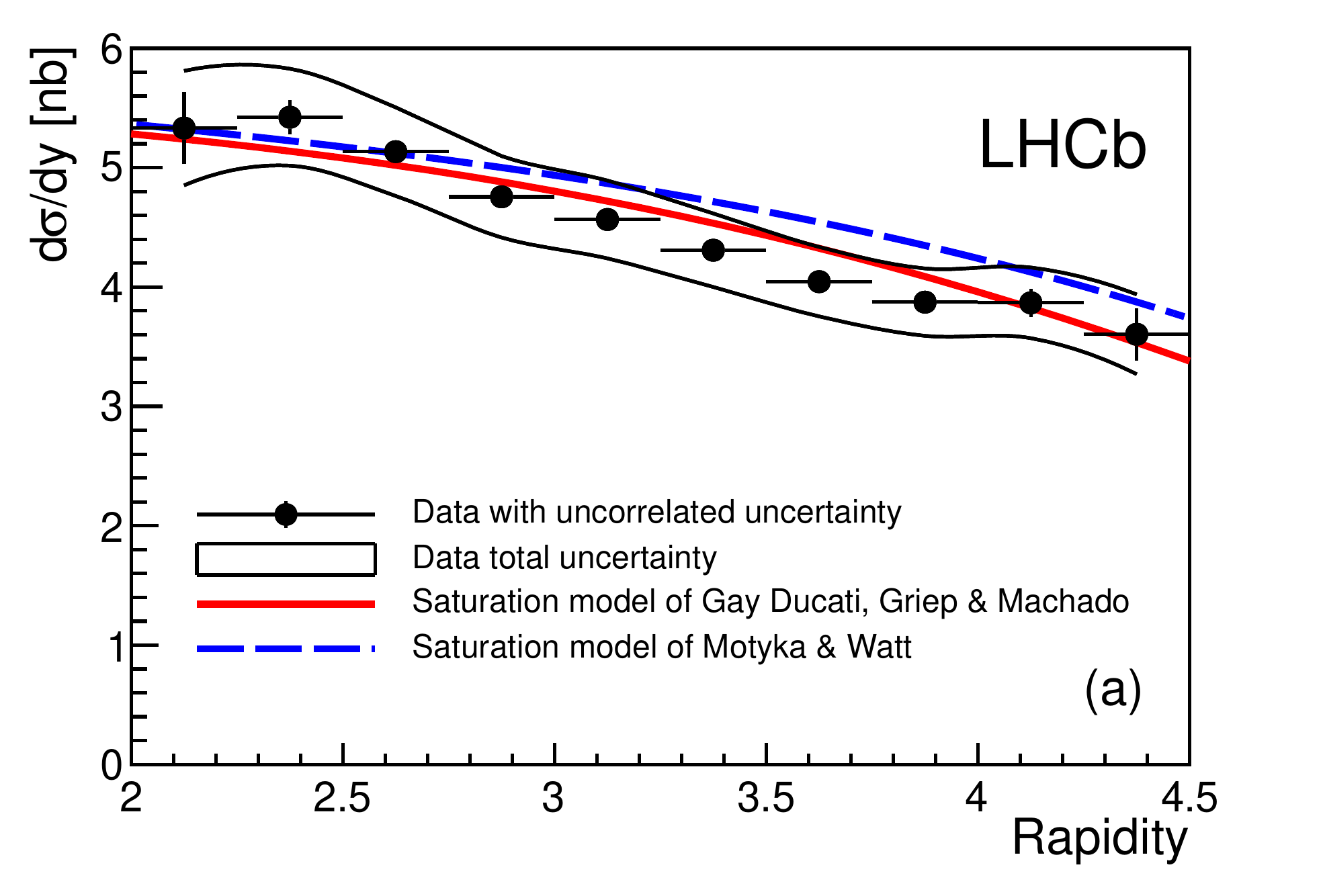}
        \includegraphics[width=7.5cm]{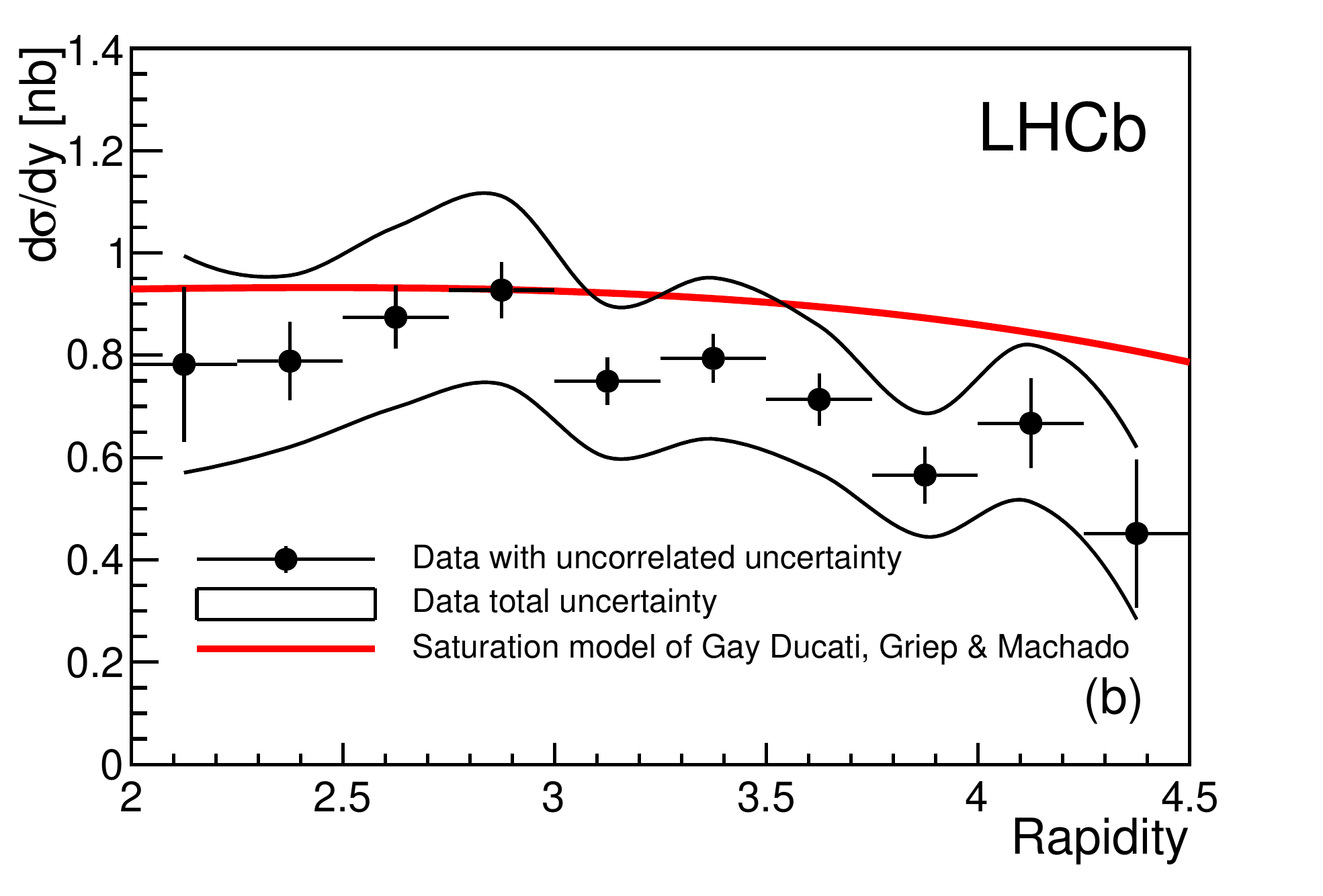}
    \vspace*{-0.8cm}
  \end{center}
  \caption{
    \small 
    Differential distribution for  (a)  \jpsi and (b) \psitwos compared to the predictions of Gay Ducati, Griep and Machado~\cite{gay} and Motyka and Watt~\cite{watt}, which include saturation effects.
        The points are data where the error bars
    indicate the uncorrelated uncertainties.  
    The band indicates the total uncertainty, most of which is correlated between bins.
          }
  \label{fig:gay}
\end{figure}

In addition to higher order effects being capable of explaining the deviation from a pure power-law
behaviour, saturation effects may be important.
Figure~\ref{fig:gay}(a) compares the \jpsi differential distribution to predictions by
Motyka and Watt~\cite{watt} and  Gay Ducati, Griep and Machado~\cite{gay}, 
that both include saturation effects and have a precision of  $10-15$\%.
A rapidity gap survival factor of $r(y)=0.85-0.1|y|/3$ has been applied to the former
while the latter assumes $r(y)=0.8$.
Both predictions use a Weizs\"acker-Williams
approximation to describe the photon flux. 
The agreement with the LHCb data is good.
Figure~\ref{fig:gay}(b) compares the \psitwos differential distribution to the 
prediction of
Gay Ducati, Griep and Machado.
Good agreement with the data is again observed.

\section{Conclusions}
The differential and integrated cross-section times branching fraction 
for \jpsi and \psitwos mesons decaying to two muons, both with $2.0<\eta<4.5$, have been measured. 
The results of this analysis are consistent with the previously published LHCb analysis, 
which used data taken in 2010,
but have a significantly improved precision, as well as a more extensive use of data-driven
techniques to estimate systematic sources.
An increase in luminosity, lower pile-up running conditions,  
as well as improvements in the trigger lead 
to roughly 40 times as many events in the 2011 data-taking period.  
The integrated cross-section measurements have an overall uncertainty that is a factor two better;
they are limited by the theoretical modelling of the inelastic background
for the \jpsi analysis
and by the statistical precision with which the background is determined for the \psitwos analysis.
The cross-section is presented differentially for the \psitwos for the first time.
Although the total uncertainty for the \jpsi differential distribution is 7\% per rapidity bin,
most of this is correlated bin-to-bin;
the uncorrelated uncertainty is typically 1.5\%. 
Thus the overall shape of the 
differential distribution is rather well determined and this improves the ability of the data
to distinguish between different theoretical models.

The integrated cross-sections are in good agreement with several theoretical estimates.
The differential \jpsi and \psitwos cross-section both agree better with the NLO rather than LO
predictions of
\cite{jones}.
The result has also been compared to two models that include saturation effects~\cite{watt,gay};
in both cases, good agreement is observed.
It is also worth noting that the $t$-dependence for LHCb and HERA results are in agreement
with Regge theory.

%% file: acknowledgements.tex
\section*{Acknowledgements}
 
\noindent 
We thank Lucian Harland-Lang and Stephen Jones for many helpful discussions and clarifications.
We express our gratitude to our colleagues in the CERN
accelerator departments for the excellent performance of the LHC. We
thank the technical and administrative staff at the LHCb
institutes. We acknowledge support from CERN and from the national
agencies: CAPES, CNPq, FAPERJ and FINEP (Brazil); NSFC (China);
CNRS/IN2P3 and Region Auvergne (France); BMBF, DFG, HGF and MPG
(Germany); SFI (Ireland); INFN (Italy); FOM and NWO (The Netherlands);
SCSR (Poland); MEN/IFA (Romania); MinES, Rosatom, RFBR and NRC
``Kurchatov Institute'' (Russia); MinECo, XuntaGal and GENCAT (Spain);
SNSF and SER (Switzerland); NAS Ukraine (Ukraine); STFC (United
Kingdom); NSF (USA). We also acknowledge the support received from the
ERC under FP7. The Tier1 computing centres are supported by IN2P3
(France), KIT and BMBF (Germany), INFN (Italy), NWO and SURF (The
Netherlands), PIC (Spain), GridPP (United Kingdom). We are thankful
for the computing resources put at our disposal by
Yandex LLC (Russia), as well as to the communities behind the multiple open
source software packages that we depend on.